\documentclass[pdflatex,sn-mathphys-num,referee]{sn-jnl}

\usepackage{mathtools}
\usepackage{slashed}
\usepackage{tensor}
\usepackage{extarrows}
\usepackage{amssymb}
\usepackage[T1]{fontenc}

\newcommand{\te}{\mathrm{e}}
\newcommand{\ti}{\mathrm{i}}
\newcommand{\td}{\mathrm{d}}
\newcommand{\erf}{\mathrm{Erf}}
\usepackage[normalem]{ulem}

\begin{document}

    \title[Near-Horizon Deformation of Metric and the Black Hole Instability]{Near-Horizon Deformation of Metric and the Black Hole Instability}
    
    \author[1]{\fnm{Shi-Jie} \sur{Ma}}
    \author[2]{\fnm{Zhan-Feng} \sur{Mai}}
    \author*[1]{\fnm{Run-Qiu} \sur{Yang}}\email{aqiu@tju.edu.cn}
    
    \affil[1]{\orgdiv{Center for Joint Quantum Studies and Department of Physics, School of Science},
    \orgname{Tianjin University}, \street{Yaguan Road 135}, \city{Tianjin},
    \postcode{300350}, \country{P. R. China}}
    
    \affil[2]{\orgdiv{Guangxi Key Laboratory for Relativistic Astrophysics, School of Physical Science and Technology},
    \orgname{Guangxi University}, \city{Nanning}, \postcode{530004}, \country{P. R. China}}
    
    \abstract{Recent time-domain analyses suggest that black hole stability may be sensitive to localized near-horizon geometric deformations, while the underlying spectral mechanism remain unclear. In this work, after verifying the instability of the model in the time domain, we systematically investigate quasi-normal mode spectra under static localized non-positive perturbations within a frequency-domain framework. We find that such deformations generically induce a new purely imaginary mode. As the deformation approaches the horizon, the imaginary part of this mode increases and eventually enters the upper half complex-frequency plane, signaling the onset of black hole instability. Numerical results reveal clear scaling relations between the critical distance for instability and the deformation strength. We further derive rigorous proofs for our discoveries in frequency domain. These results demonstrate that black hole stability under long scale is conditionally sensitive to localized deformation of metric near the horizon and establish a unified spectral framework for understanding their induced instabilities.}
    
    \maketitle

	\section{Introduction}\label{sec1}

	Quasi-normal modes (QNMs), the characteristic oscillatory modes of perturbed black holes (BHs) and compact objects, dominate the ringdown stage following binary mergers~\cite{PhysRevD.93.044020}. Their complex frequencies-whose real parts determine the oscillation frequencies and imaginary parts characterize the damping timescales-encode essential information about the underlying spacetime geometry~\cite{kokkotas_quasi-normal_1999,Berti_2009,RevModPhys.83.793}. This unique feature has motivated the development of BH spectroscopy, which uses QNM spectra to test general relativity~\cite{PhysRevLett.117.091102}, examine the no-hair theorem and BH uniqueness~\cite{PhysRevD.96.104040,PhysRevLett.118.161101,PhysRevLett.123.111102}, and distinguish BHs from exotic compact objects~\cite{PhysRevLett.116.171101,PhysRevD.73.064030,PhysRevD.94.084031}.
	
	Since the 1990s, Nollert investigated the incompleteness of the QNM spectrum by approximating the Schwarzschild Regge-Wheeler (RW) potential with step functions and discovered dramatic spectral shifts under arbitrarily small perturbations~\cite{PhysRevD.53.4397}. Together with Price, he further showed that perturbed QNMs migrate to new branches with distinct asymptotic behaviors, establishing the framework of BH spectral instability~\cite{10.1063/1.532698}. This phenomenon has since attracted extensive attention~\cite{PhysRevX.11.031003,PhysRevLett.128.211102,Jaramillo_2022,PhysRevD.104.084091,PhysRevLett.128.111103,PhysRevD.107.064012}, including studies on the sensitivity of overtones to near-horizon deformations~\cite{Konoplya2022pbc,Konoplya2023hqb}. A common conclusion of these studies is that infinitesimal perturbations can induce significant spectral migration while leaving the time-domain waveform nearly unchanged~\cite{PhysRevD.101.104009,PhysRevD.106.084011,PhysRevD.110.084018,PhysRevLett.133.211401,PhysRevD.110.L121501}. In particular, Sheikh \textit{et al.}~\cite{PhysRevX.11.031003} showed that higher overtones migrate toward the real axis, with sensitivity increasing with mode order. Cardoso \textit{et al.}~\cite{PhysRevLett.128.111103} further found that moving the deformation away from the main potential barrier leads to repeated overtaking of the fundamental mode by higher overtones, eventually resulting in drastic spectral deviations. These analyses, though deform the potential in different ways, they still keep the total potential strictly nonnegative everywhere. Since no fundamental physical principle requires perturbations to be positive definite, the effect of non-positive perturbations on QNM spectra remains largely unexplored.
	
	Recently, Ref.~\cite{mai_numerical_2026} introduced two types of non-positive localized perturbations near the horizon of an otherwise stable BH: a negative bump and zero-mean stochastic perturbations, and found exponentially growing time-domain waveforms that signal a dynamical instability. This finding suggests that BH stability may not be robust under near-horizon non-positive perturbations. Additionally, such instability does not emerge inevitably; rather, it appears only when the deformation is placed beyond a critical distance $a_c$, which scales with the deformation strength $\epsilon$ according to distinct power laws:
	\begin{equation}\label{eq1}
		a_c\propto
		\begin{cases}
			\epsilon^{-1}, & \text{negative bump perturbations},\\
			\epsilon^{-2}, & \text{zero-mean stochastic perturbations}.
		\end{cases}
	\end{equation}
    Although Ref.~\cite{mai_numerical_2026} also reported the crossing phenomenon of the instability-related purely imaginary mode in the frequency domain, the complete spectral evolution leading to this crossing, as well as its underlying mechanism, remains unresolved. This naturally raises the following fundamental questions:
	\begin{itemize}
        \item Is the crossing phenomenon a special  case or a common phenomenon?
		\item What is the complete frequency-domain evolution of the instability-related mode prior to the onset of instability?
		\item Under what precise conditions can non-positive perturbations trigger BH instability?
		\item Can the power laws in Eq.~\eqref{eq1} be derived theoretically?
	\end{itemize}
	
	In this work, we first derive how near-horizon geometric fluctuations modify the effective potential. Based on this result, we show that stochastic deformations of the effective potential containing both positive and negative contributions arise naturally from geometric fluctuations. We then separately investigate the effects of negative bumps and stochastic perturbations on the QNM spectra of the double-$\delta$ potential, P\"{o}schl-Teller (PT) potential, and Regge-Wheeler (RW) potential in the frequency domain. Our results show that negative and stochastic perturbations not only drive high-overtone modes rapidly toward the real axis, as observed for positive perturbations, but also generate an additional purely imaginary mode that encodes information about BH instability. As the deformation moves closer to the horizon, this mode travels along the imaginary axis from the lower half of the complex plane, corresponding to stable configurations, to the upper half plane, signaling the onset of instability. By drawing an analogy with the Schr\"{o}dinger equation, we further demonstrate theoretically that arbitrarily small negative and stochastic localized perturbations can always trigger BH instability, provided they are sufficiently close to the horizon. We finally derive Eq.~\eqref{eq1} analytically and prove that the mode carrying information about BH instability must necessarily be purely imaginary.
	
	This paper is organized as follows. In Sec.~\ref{sec2}, we show how near-horizon fluctuations can be modeled as stochastic perturbations in the tortoise coordinate and how they induce effective perturbations in the potential. In Sec.~\ref{sec3}, we investigate deterministic negative perturbations in three representative potential models and demonstrate how instability can arise. This can be regarded as the counterpart of positive deformation on the effective potentials in previous literatures and also a toy model to understand the effect of negative parts in the stochastic deformation of effective potentials. In Sec.~\ref{sec4}, we extend the analysis to stochastic perturbations and show that even perturbations with vanishing spatial averages can trigger instability. In Sec.~\ref{sec5}, we present a theoretical analysis and establish a rigorous instability criterion for stochastic perturbations. Finally, we summarize our conclusions in Sec.~\ref{sec6}.

	\section{From Near-Horizon Fluctuations to Effective Potential Deformations}\label{sec2}
	
	In this section, we will argue that the local stochastic deformation on the effective potential will raise naturally if the near-horizon geometry could be continuously deformed by some stochastic factors. In realistic astrophysical settings, BHs interact with surrounding matter and fields. Such effects, together with classical and quantum fluctuations, introduce small perturbations to the spacetime geometry. Strictly speaking, such perturbations deform the metric in spatial coordinates as well as in temporal coordinate and so we have to consider a time-dependent geometry. To simplify our discussion, we focus on the case that perturbations are static in time but have local stochastic distributions in spatial coordinates. In such setup, we consider a static, spherically symmetric, and asymptotically flat BH spacetime written in tortoise coordinate $x$,
	\begin{equation}\label{metric}
		\td s^2 = h(x)\left(-\td t^2 + \td x^2\right) + r(x)^2 \td\Omega^2.
	\end{equation}
	In this coordinate system, the event horizon corresponds to the limit $x \to -\infty$, where the radial function satisfies $r(-\infty) = r_h$ and the metric function vanishes, $h(-\infty) = 0$.
	
	Due to asymptotic flatness, the spacetime approaches the Minkowski form in the far-field region. In this limit, one has $r(x) \sim x$ as $x \to +\infty$, and the metric function can be approximated as $h(x) \simeq 1 - \frac{2M}{x}$. Meanwhile, the radial function satisfies the boundary condition $r'(x) \to 1$ as $x \to +\infty$, ensuring consistency with the flat spacetime limit.
	
	To investigate the stability of the spacetime, we consider the simplest model, namely, the stability under an $s$-wave scalar perturbation $\Psi$ treated as a test field. Following the standard derivation of scalar perturbations in a static spherically symmetric spacetime~\cite{PhysRevD.101.104009}, the Klein--Gordon equation $\Box\Psi=0$ can be written as
	\begin{equation}\label{KGet}
		\left(\partial_x^2 - \partial_t^2 - V_{\text{eff}}\right)\tilde\psi = 0,
	\end{equation}
	where $\tilde\psi = r\Psi$, and the effective potential is given by the RW form $V_{\text{eff}} = r''/r$. Assuming the harmonic time dependence $\tilde\psi(t,x)=\psi(x)\te^{-\ti\omega t}$, the above equation becomes
    \begin{equation}\label{KGe}
		\left(\partial_x^2 + \omega^2 - V_{\text{eff}}\right)\psi = 0,
	\end{equation}
    where $\omega$ is the frequency of $\psi$. Eq.~\eqref{KGe} serves as the starting point for the frequency-domain calculations presented in the following sections, while the time-domain evolutions are obtained by solving Eq.~\eqref{KGet}.
	
	We now sperate the metric function $h$ and $r$ into two parts
	\begin{equation}\label{eqr}
		h=h_0+\delta h,\quad r = r_0 + \delta r.
	\end{equation}
	Here $h_0$ and $r_0$ denote the unperturbed metric functions, where are given by the ``idealized model'' and do not include stochastic factors around the BH. The $\delta h$ and $\delta r$ then represents a stochastic deformation induced by environmental effects and unavoidable fluctuations. The deformation is localized around $x = a$, with support in the interval $x_- < x < x_+$, i.e.
    \begin{equation}\label{deltahr}
      \delta h,~\mathrm{and}~\delta r\begin{cases}
      \neq0,&x\in(x_-,x_+),\\
      =0,&x\notin(x_-,x_+).
      \end{cases}
    \end{equation}
    Since the effective potential considered in this work is given by $V_{\mathrm{eff}}=r''/r$, the perturbation $\delta h$ does not contribute to $V_{\mathrm{eff}}$. Therefore, the subsequent analysis only involves the deformation $\delta r$. We assume that the deformation has a vanishing spatial average,
	\begin{equation}\label{intdeltar}
		\overline{\delta r} = \frac{1}{x_+ - x_-} \int_{x_-}^{x_+} \delta r \, \td x = 0.
	\end{equation}
	We further impose smoothness conditions at the boundaries,
	\begin{equation}\label{smooth}
		\left. \frac{\td^n \delta r}{\td x^n} \right|_{x=x_-} =
		\left. \frac{\td^n \delta r}{\td x^n} \right|_{x=x_+} = 0, \quad n=0,1,2.
	\end{equation}
	
    Expanding the effective potential, we obtain
	\begin{equation}\label{eqV}
		V_{\text{eff}} = \frac{r_0''}{r_0} + \frac{\delta r''}{r_0} - \frac{r_0''}{r_0^2}\,\delta r + \mathcal{O}(\delta r^2).
	\end{equation}
	Retaining terms up to first order, we decompose
	\begin{equation}\label{deltaV}
		V_0 = \frac{r_0''}{r_0}, \qquad \delta V = \frac{\delta r''}{r_0} - \frac{r_0''}{r_0} \cdot \frac{\delta r}{r_0},
	\end{equation}
	where $V_0$ is the unperturbed effective potential, and $\delta V$ is the stochastic deformation induced by $\delta r$.

    We here consider the ``near-horizon'' local deformation on metric, so we assume that $x_-$ and $x_+$ approaches into $-\infty$. Keep in mind that $x$ is the tortoise coordinate, above assumption means that $r_0(x)\rightarrow r_h$ and $r_0''(x)\rightarrow0$. Therefore,
    \begin{equation}\label{intdeltaV}
		\delta V = \frac{\delta r''}{r_h}\neq0,\quad \overline{\delta V}=
		\frac{1}{x_{+} - x_{-}}\frac1{r_h}\int_{x_{-}}^{x_{+}}\delta r'' \td x=0.
	\end{equation}
	This result implies that localized perturbations with vanishing average can induce deformation of effective potential with zero average, so we should take both positive and negative deformations of the effective potential into account. To fully understand how near-horizon fluctuations affect BH stability, it is necessary to consider both negative and positive perturbations.

	\section{Instability Induced by Deterministic Negative Bumps}\label{sec3}
	As discussed in the previous section, a complete understanding of the impact of near-horizon fluctuations on the QNM spectrum and BH stability requires taking negative perturbations into account. In this section, we model such effects by introducing a localized static negative bump in the effective potential. Although this setup does not fully capture realistic stochastic perturbations, it serves as a simple toy model that captures their essential features. For convenience, we write the effective potential as
	\begin{equation}
		V_{\text{eff}} = V_b + \epsilon V_p,
	\end{equation}
	where $V_b$ is the background potential, $V_p$ is a localized deformation centered at $x=a$, and $\epsilon$ denotes its strength. The corresponding boundary conditions are
	\begin{equation}\label{boundary}
		\psi(+\infty)\to \te^{\ti\omega x},\qquad \psi(-\infty)\to \te^{-\ti\omega x}.
	\end{equation}

	Intuitively, one might expect that a negative deformation $V_p$ can support a trapped state, potentially leading to QNMs with positive imaginary parts and thus rendering the BH unstable. This is indeed the case when $V_p$ is considered in isolation. However, in the presence of the background potential $V_b$, the interplay between the two potentials becomes nontrivial. When $V_p$ is sufficiently close to $V_b$, their competition can prevent the formation of bound states, as will be demonstrated below. 
	
	\subsection{Double-\texorpdfstring{$\delta$}{delta} Potential: Emergence of Instability from a Negative \texorpdfstring{$\delta$}{delta} Perturbation}\label{sec3subnd}
    
	To facilitate analytical treatment, we model both $V_b$ and $V_p$ using $\delta$-potentials as a toy model. A single $\delta$-potential with strength $V_0$ supports only one QNM with frequency $-\ti V_0/2$. Therefore, we adopt a double-$\delta$ potential, which admits multiple QNMs,
	\begin{equation}\label{V_negative_delta}
		V_{b}(x)=\delta(x-b_0/2)+\delta(x+b_0/2),~
		V_{p}(x,a)=-\delta(x-a).
	\end{equation}
	where $b_0$ is the distance between the two $\delta$-potentials.
	
	\begin{figure*}[htbp]
		\centering
		\includegraphics[width=0.99\textwidth]{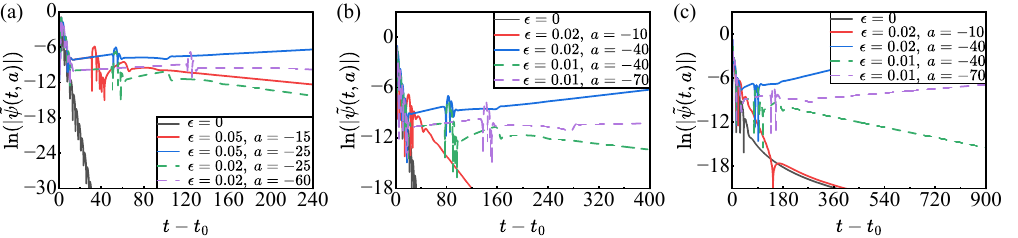}
		\caption{Time evolution of the deformation field for different potential models with localized negative bumps. Panels (a)-(c) correspond to the double-$\delta$ potential, PT potential, and RW potential, respectively. Here $t_0$ is defined as the first time at which $|\tilde\psi(t,a)|$ is at its local maximum value.}\label{fignt}
	\end{figure*}
	We first verify whether the model exhibits the growing time evolution curve that indicates the BH instability reported in Ref.~\cite{mai_numerical_2026} in the time domain. We investigate the time evolution of a Gaussian wave packet with initial conditions $\tilde\psi(0,x)=\te^{-8x^2}$ and $\td\tilde\psi(0,x)/\td t=0$ for different values of $\epsilon$. The Gaussian wave packet is adopted as a localized initial perturbation with a broad frequency spectrum, allowing the characteristic modes of the system to be efficiently excited. Unless otherwise specified, the same initial perturbation is used throughout all time-domain simulations in this work. More generally, the specific Gaussian profile is not essential, and any sufficiently smooth and rapidly decaying localized perturbation can be used as the initial data.
    
    We choose $b_0=1$ for the double-$\delta$ potential. Fig.~\ref{fignt}(a) shows the results for perturbations away from $V_b$ with two strengths: $\epsilon=0.05$ (solid colored line) and $\epsilon=0.02$ (dashed colored line), with the unperturbed case (solid black line) as a control group. The unperturbed case exhibits oscillatory exponential decay, where the fundamental frequency $\omega_0$ is complex with $\Im(\omega_0)<0$.When the bump is located near the main potential barrier, the curve enters exponential decay accompanied by echo oscillations after several oscillations, indicating that the BH remains stable, although a purely imaginary mode $\varpi_0$ has already overtaken the original fundamental mode and become dominant. When the bump is located closer to the horizon (This corresponds that $a$ is sufficiently negative), after several initial oscillations, the curve exhibits an overall exponential growth with superimposed echo oscillations, indicating the onset of instability of the BH. These echoes originate from repeated scattering between the main potential barrier and the localized perturbation. In particular, the time interval between successive echoes is approximately $2|a|$, corresponding to the round-trip propagation time between the two scattering structures, as expected for echo signals~\cite{PhysRevD.106.084011}.
    This behavior is consistent with the existence of a critical distance $a_c$ at which the BH transitions from stable to unstable. The corresponding results for the PT and RW potentials shown in Figs.~\ref{fignt}(b) and (c), respectively, are discussed in Sec.~\ref{sec3subnp}
	
	\begin{figure*}[htbp]
		\centering
		\includegraphics[width=0.99\textwidth]{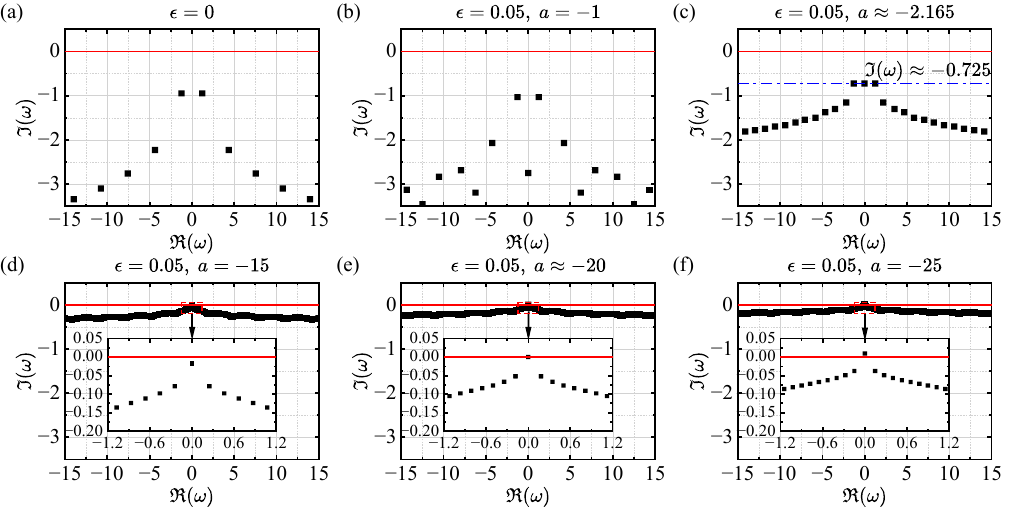}
		\caption{QNM spectrum of the double-$\delta$ potential with a negative $\delta$-bump for $\epsilon=0.05$ and different values of $a$. }\label{figndf}
	\end{figure*}
	
	We now investigate what occurs in the frequency domain as the bump is located progressively closer to the horizon. Recently, Ref.~\cite{mai_numerical_2026} reported that an instability-related purely imaginary mode crosses into the upper half of the complex-frequency plane in a simplified toy model. Here, we further investigate the spectral evolution of this instability-related branch in the present model. Under the QNM boundary conditions in Eq.~\eqref{boundary}, Eq.~\eqref{KGe} constitutes an eigenvalue problem for the complex frequency $\omega$, whose discrete eigenvalues correspond to the QNM frequencies. We determine these frequencies by direct numerical integration. For a trial complex frequency $\omega$, two independent solutions satisfying the QNM boundary conditions at the left and right asymptotic boundaries are integrated toward a matching point near the peak of the main potential barrier. The QNM frequencies are then determined by requiring the two solutions to match, equivalently, by searching for the zeros of their Wronskian at the matching point. This direct integration and matching procedure is used throughout the frequency-domain calculations in this work. For the double-$\delta$ potential considered here, the detailed implementation is presented in Appendix~\ref{App_delta} For this purpose, we plot the frequency spectra for $\epsilon=0$ and various values of $|a|$ in Fig.~\ref{figndf}, with the unperturbed case again serving as a reference.
	\begin{itemize}
		\item[(1)] In the unperturbed case (Fig.~\ref{figndf}(a)), all QNM frequencies lie in the lower half-plane and form symmetric branches with respect to the imaginary axis, with nonzero real parts. {Its corresponding time-evolution curve is given by the case with $\epsilon=0$ in Fig.~\ref{fignt}(a).}
		
		\item[(2)] When the negative bump is located near the main potential barrier (Fig.~\ref{figndf}(b), $|a|=1$), the original spectral branch is only slightly perturbed, while a new branch emerges below it. A purely imaginary mode $\varpi_0$ also appears and possesses the largest imaginary part within this new branch. Although the spectrum of high overtones is affected dramatically, $\varpi_0$ has not yet overtaken the original fundamental mode $\omega_0$ and the fundamental mode $\omega_0$ is only changed slightly at this stage. Its time-evolution curve should behave similarly to the unperturbed case. It should be noted that the appearance of this new branch depends on the perturbation strength. For the same deformation position $a$, our numerical results show that decreasing the magnitude of the negative perturbation shifts the new branch toward increasingly negative values of $\Im(\omega)$. Due to finite numerical precision, the branch cannot be resolved for arbitrarily weak perturbations; nevertheless, this systematic trend suggests that a new branch is induced by any nonzero negative perturbation, while for sufficiently weak perturbations it lies too far down in the complex-frequency plane to be numerically resolved.

        \item[(3)] As the bump is located closer to the horizon (Fig.~\ref{figndf}(c), $a\approx -2.165$), both branches migrate toward the real axis, with the new branch migrating faster and nearly merging with the original branch. Meanwhile, the high-overtone spectral points become substantially more numerous. These two effects make an unambiguous distinction between the two branches difficult. At this point, $\Im(\varpi_0)=\Im(\omega_0)$, marking the overtaking of the original fundamental mode.
		
		\item[(4)] When the bump is located even closer to the horizon (Fig.~\ref{figndf}(d)), $\varpi_0$ overtakes the original fundamental mode and becomes the dominant mode. Although $\Im(\varpi_0)<0$ still holds, the late-time behavior becomes purely exponentially decaying without oscillations, explaining the absence of oscillatory behavior in the corresponding time evolution. {The corresponding time-evolution curves are illustrated by the cases $\epsilon=-0.05,\,a=-15$ and $\epsilon=-0.02,\,a=-25$ in Fig.~\ref{fignt}(a).}
		
		\item[(5)] As the bump approaches the horizon further (Fig.~\ref{figndf}(e)), $\varpi_0$ continues to move along the imaginary axis toward the origin of the complex frequency plane, and the BH approaches the threshold of instability. A similar behavior of the instability-related purely imaginary mode near the instability threshold was also observed in Ref.~\cite{mai_numerical_2026}.
		
		\item[(6)] When the bump is sufficiently close to the horizon (Fig.~\ref{figndf}(f)), $\varpi_0$ crosses into the upper half-plane, signaling the onset of BH instability. Since $\varpi_0$ is purely imaginary with $\Im(\varpi_0)>0$, it corresponds to exponential growth in the late-time evolution. The corresponding time-evolution curves are illustrated by the cases $\epsilon=-0.05,\,a=-25$ and $\epsilon=-0.02,\,a=-60$ in Fig.~\ref{fignt}(a).
	\end{itemize}
    Owing to the symmetric two-branch structure of the double-$\delta$ potential, the present model further allows us to follow the preceding evolution of this instability-related branch, including its appearance and overtaking of the original fundamental mode, until it eventually crosses into the upper half of the complex-frequency plane at the critical distance $a_c$, signaling the onset of BH instability.

	\begin{figure*}[htbp]
		\centering
		\includegraphics[width=0.99\textwidth]{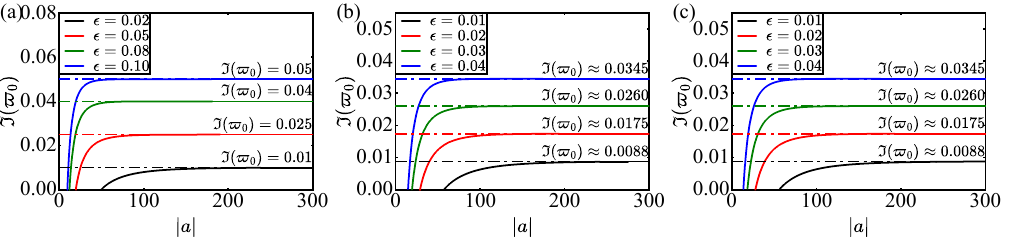}
		\caption{Imaginary part of the dominant mode $\varpi_0$ as a function of $|a|$ for different deformation strengths $\epsilon$. Panels (a)-(c) correspond to the double-$\delta$, PT, and RW potentials, respectively.}\label{figna}
	\end{figure*}

	To investigate whether the frequency of $\varpi_0$ increases and approaches a finite value in the limit where the bump approaches the event horizon, we plot the curves of $\Im(\varpi_{0})$ varying with $|a|$ for different values of $\epsilon$ in Fig.~\ref{figna}(a). The figure shows that for all different $\epsilon$ values, $\Im(\varpi_{0})$ approaches $\ti\epsilon/2$ (i.e., the fundamental mode of the isolated $\epsilon V_p$) as $a\to-\infty$, indicating that when $V_p$ is sufficiently close to the horizon, the late-time behavior of the BH is completely dominated by $V_p$. Meanwhile, we also observe that there is an inverse relationship between $|a_c|$ and $\epsilon$, i.e., $|a_c|\approx\epsilon^{-1}$ (for example, in Fig.~\ref{figndf}(e), we observed that $|a_c|\approx20=\epsilon^{-1}$ with $\epsilon=0.05$), which is consistent with the observation reported in Ref.~\cite{mai_numerical_2026}. The corresponding results for the PT and RW potentials shown in Figs.~\ref{figna}(b) and (c), respectively, are discussed in Sec.~\ref{sec3subnp}.

	\subsection{Instability in Smooth Potentials}\label{sec3subnp}
	In the previous subsection, the double-$\delta$ model was used to expose several characteristic features of the frequency-domain spectrum under negative perturbations. However, as a highly idealized setup, it does not by itself establish a general mechanism, since such a mechanism must be supported by behavior that is robust across different potentials. We therefore turn to more realistic and smooth potentials to examine whether these features persist beyond the toy model. Specifically, we consider the PT potential and the RW potential. As we will show, the same spectral features appear in both cases, indicating that they are not model-specific but reflect a universal mechanism.

	We first consider the PT potential, which provides a smooth and analytically tractable approximation to BH effective potentials.
	We introduce a localized negative Gaussian bump,
	\begin{equation}\label{V_negative_pt}
		V_{b}=\frac{1}{\cosh^2 (x)},\qquad
		V_{p}=-\te^{-(x-a)^2}.
	\end{equation}
	
	The perturbed PT potential exhibits behavior closely analogous to that of the double-$\delta$ model. In the time domain (Fig.~\ref{fignt}(b)), the evolution shows exponential decay when the bump is located near the main barrier, while exponential growth sets in once the bump is sufficiently close to the horizon, signaling the onset of instability. This indicates that negative bumps destabilize the PT system, and as in the double-$\delta$ model, a critical separation $a_c$ separates stable and unstable regimes.

	\begin{figure*}[htbp]
		\centering
		\includegraphics[width=0.99\textwidth]{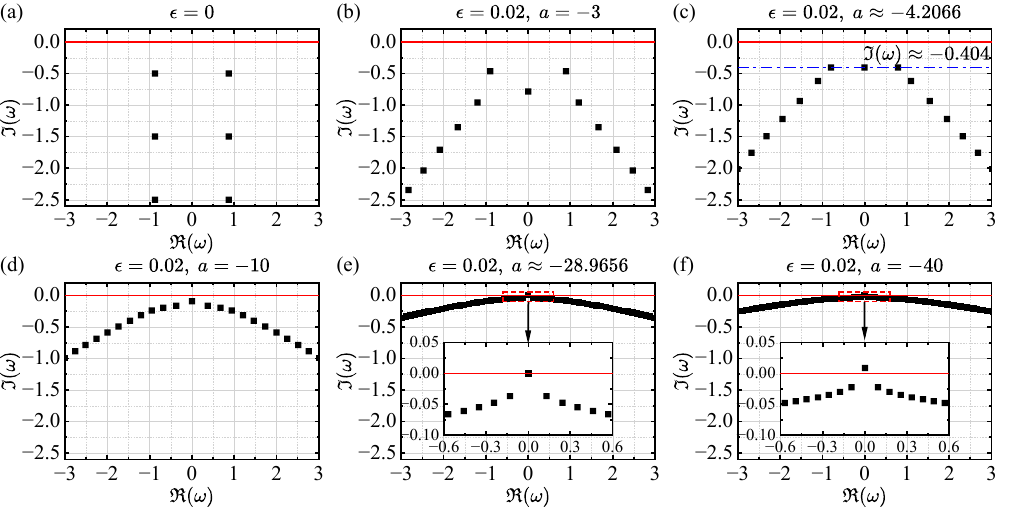}
		\caption{QNM spectra of the PT potential with a localized Gaussian bump for $\epsilon=0.02$ and different values of $a$.}\label{fignpf}
	\end{figure*}
    In the frequency domain (Figs.~\ref{fignpf} and \ref{figna}(b); see Appendix~\ref{App_pt} for the QNM calculation method for the perturbed PT potential), many features are consistent with the double-$\delta$ case. A pure imaginary mode $\varpi_0$ appears again on the negative imaginary axis, while the fundamental mode only undergoes a slight shift relative to the unperturbed case; correspondingly, the time-evolution curve exhibits oscillatory decay similar to the unperturbed one (Fig.~\ref{fignt}(b), $\epsilon=0$). As the negative bump is pushed toward the horizon ($x \to -\infty$), higher overtones rapidly approach the real axis, $\varpi_0$ migrates along the imaginary axis and overtakes the original fundamental mode, for which the time-evolution curve displays exponential decay (Fig.~\ref{fignt}(b), $\epsilon=0.02,\,a=-10$ and $\epsilon=0.01,\,a=-40$). It subsequently crosses into the upper half-plane at the critical distance $a_c$, signaling the onset of instability, {where the time-evolution curve shows exponential growth (Fig.~\ref{fignt}(b), $\epsilon=0.02,\,a=-40$ and $\epsilon=0.01,\,a=-70$)}. In this asymptotic limit, $\varpi_0$ approaches the fundamental frequency of the isolated potential $\epsilon V_p$, obeying the same scaling relation $a_c \propto \epsilon^{-1}$.

	We now study the effect of localized negative perturbations on the stability of Schwarzschild BH. We take the RW potential with $\ell=s=0$ as the background potential $V_b$ and describe the deformation in the form of a Gaussian distribution.
	\begin{equation}\label{V_random_rw}
		V_{b}=\frac{1}{r^3}-\frac{1}{r^4},\qquad V_{p}=-\te^{-(x-a)^2}.
	\end{equation}
	The time evolution of $\tilde\psi$ is shown in Fig.~\ref{fignt}(c), which is somewhat different from those of the previous several models. In the unperturbed case, the waveform exhibits the characteristic power-law tail. For cases where the bump is sufficiently close to the main barrier, the signal undergoes several damped oscillations, followed by exponential decay, and eventually transitions to a power-law tail. For cases where the bump is sufficiently close to the horizon, the waveform instead evolves into exponential growth after several oscillations and does not return to the power-law regime, indicating that the BH does not return to a stable state represented by the power-law tail after becoming unstable. Echo-like features are also present in the perturbed cases. However, the additional oscillation observed during the transition from exponential decay to the power-law tail does not satisfy the characteristic relation $t - t_0 = 2n|a|$ where $n \in \mathbb{Z}^{+}$. It should therefore not be attributed to the echo mechanism, but rather to the transition between the QNM-dominated regime and the late-time power-law tail. It is just that it was obscured by the oscillation of the fundamental mode in the unperturbed curve.

	\begin{figure*}[htbp]
		\centering
		\includegraphics[width=0.66\textwidth]{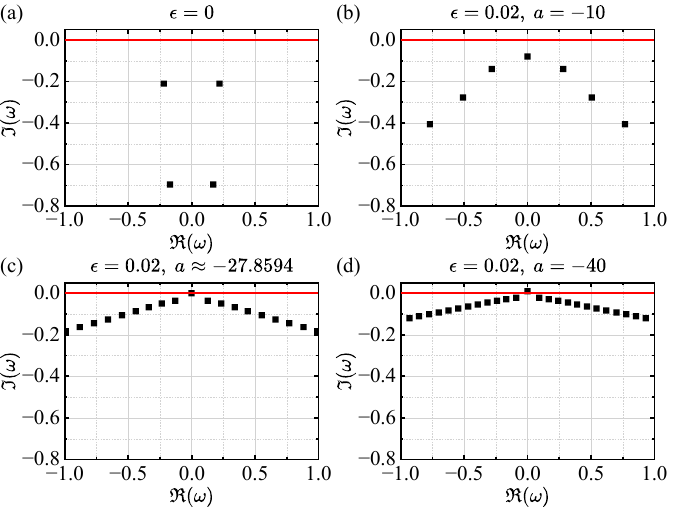}
		\caption{QNM spectra of the RW potential with negative Gaussian bump for $\epsilon=0.02$ and different values of $a$.}\label{fignrf}
	\end{figure*}
	
	Although our numerical method becomes less reliable for small $|a|$ (e.g., $|a|\approx 7$, with detailed explanations provided in Appendix~\ref{App_rw}), we restrict ourselves to the trustworthy regime in which the frequency-domain behavior (Figs.~\ref{fignrf}, \ref{figna}(c)) can be reliably extracted.

	In this regime, the RW potential exhibits a strong consistency with the PT case. The spectral evolution follows the same overall pattern: high overtones rapidly migrate toward the real axis as the negative bump approaches the horizon, while a single mode $\varpi_0$, which carries the instability information, emerges in the complex plane and crosses the real axis at $a=a_c$, signaling the onset of BH instability. In the asymptotic limit where the bump approaches the horizon, $\varpi_0$ further approaches the fundamental frequency of the isolated $\epsilon V_p$.
	
	A minor deviation is that no clear overtake phenomenon is observed in the RW case; however, within the reliable regime $\varpi_0$ already behaves as the fundamental mode, suggesting that the overtake process may have occurred outside the numerically accessible region rather than being physically absent.

	Beyond this difference, a direct comparison between the two systems with the same value of $\epsilon$ shows that the $\Im(\varpi_0)$---$a$ dependence in the unstable regime is nearly identical, sharing the same asymptotic frequency and comparable critical value $a_c$. These results further suggest that the critical separation $a_c$, in addition to satisfying Eq.~\eqref{eq1}, may also depend on the detailed shape of the negative bump.
	
	Let us make a short summary on this section. In many existing studies, localized perturbations are typically assumed to be positive. However, there is no fundamental requirement for such perturbations to be positive. In realistic scenarios, near-horizon classical or quantum fluctuations may generate deformations with vanishing average but locally negative contributions. Negative perturbations may also arise under suitable energy conditions~\cite{mai_numerical_2026}, from tachyonic fields~\cite{PhysRevD.80.127502,PhysRevD.108.104020}, or in modified gravity theories such as Einstein-Gauss-Bonnet gravity~\cite{cao_stability_2025}. Our results show that, if a small negative potential appears sufficiently close to the horizon, it may turnover the stability of a stable BH.
	
	\section{Instability from Stochastic Perturbations}\label{sec4}
    
	In the previous section, we isolated the negative component of the near-horizon from localized stochastic perturbations and modeled it as a localized negative deformation to analyze its impact on BH stability. It is clear that localized negative perturbations can indeed lead to BH instability. While such perturbations can indeed trigger BH instability, they do not necessarily do so; instead, there exists a critical distance that is inversely proportional to the deformation strength $\epsilon$. We now proceed to consider the impact of full near-horizon localized stochastic perturbations on BH stability.

    Due to the complexity of near-horizon stochastic fluctuations, their full physical description is not directly suitable for numerical implementation. We therefore introduce a simplified zero-mean localized perturbation model given by
    \begin{equation}
        V_p(x) = (x-a)\te^{-(x-a)^2},
    \end{equation}
    as an effective numerical representation of stochastic fluctuations. Though this is seemingly not a ``good'' model for ``stochastic local potential'', we will theoretically prove in next section that the key of a stochastic potential that triggers the instability is that it is local potential with zero-mean. All local potentials of zero-mean will exhibit similar behavior and their specific forms do not change the core conclusions of the numerical analysis in this section, which justifies our scheme of simplifying stochastic fluctuation perturbations using the above expression. In this section, we consider both the PT and RW potentials as background models, and analyze the effects of stochastic perturbations on the QNM spectrum and BH stability from both time-domain and frequency-domain perspectives.

	\begin{figure}[htbp]
		\centering
		\includegraphics[width=0.66\textwidth]{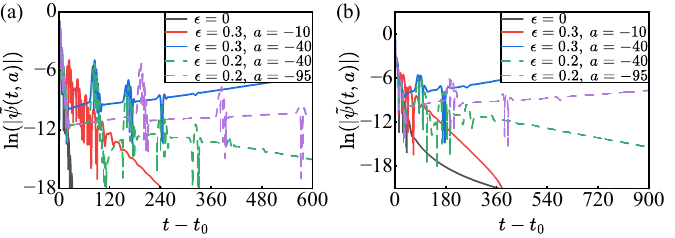}
		\caption{Time evolution of the deformation field for different potential models with localized stochastic perturbations. Panels (a) and (b) correspond to the PT potential, and RW potential, respectively. Here $t_0$ is defined as the first time at which $|\tilde\psi(t,a)|$ is at its local maximum value.}\label{figrt}
	\end{figure}
\begin{figure*}[htbp]
		\centering
		\includegraphics[width=0.99\textwidth]{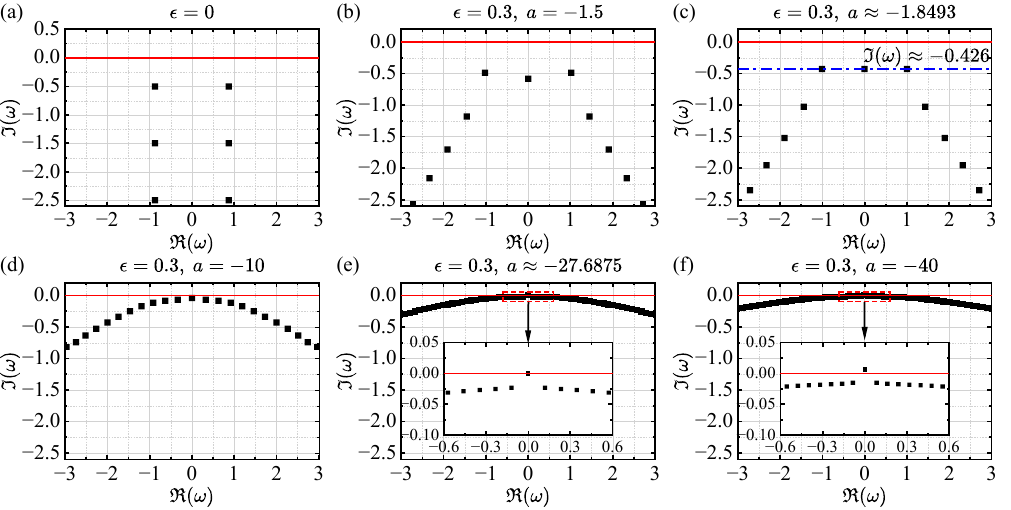}
		\caption{QNM spectra of the PT potential with a localized stochastic deformation for $\epsilon=0.3$ and different values of $a$.}\label{figrpf}
	\end{figure*}
\begin{figure*}[htbp]
		\centering
		\includegraphics[width=0.66\textwidth]{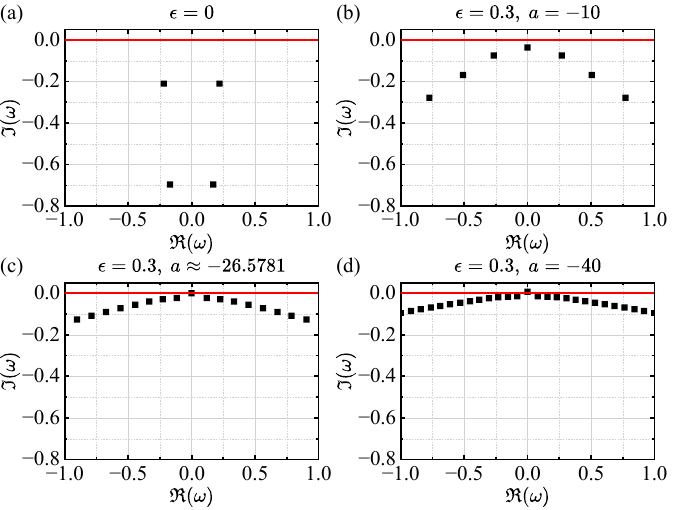}
		\caption{QNM spectra of the RW potential with a localized stochastic deformation for $\epsilon=0.3$ and different values of $a$.}\label{figrrf}
	\end{figure*}
\begin{figure}[htbp]
    	\centering
    	\includegraphics[width=0.66\textwidth]{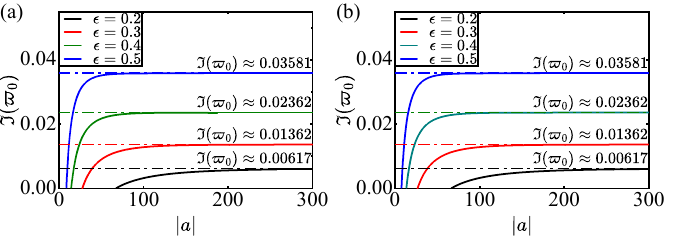}
    	\caption{Imaginary part of the dominant mode $\varpi_0$ as a function of $|a|$ for different deformation strengths $\epsilon$. Panels (a) and (b) correspond to the PT, and RW potentials, respectively. }\label{figra}
    \end{figure}

	As shown in the time domain (Fig.~\ref{figrt}) and the frequency domain (Figs.~\ref{figrpf},~\ref{figrrf}, and ~\ref{figra}), the PT and RW potentials under stochastic perturbations exhibit behavior highly consistent with that observed in the negative bump case.

	In the time domain (see Fig.~\ref{figrt}), the PT potential displays exponential decay when the deformation is located near the main barrier, and exponential growth when it is located closer to the horizon. The RW potential shows the same qualitative behavior: for perturbations near the barrier, the evolution is dominated by the ringdown phase, followed by a transition to the power-law tail, during which a single additional oscillation may appear. This oscillation occurs in the crossover between the exponential decay and late-time tail and is not associated with the echo mechanism. For perturbations located closer to the horizon, exponential growth dominates and the signal no longer returns to the tail.
	
    In the frequency domain (see Figs.~\ref{figrpf} and \ref{figrrf}), both potentials develop a purely imaginary mode $\varpi_0$ associated with instability. As the deformation is shifted toward the horizon, $\varpi_0$ migrates along the imaginary axis, overtakes the original fundamental mode, and crosses into the upper half-plane at $a=a_c$, signaling the onset of instability. In the limit where the deformation approaches the horizon ($x \to -\infty$), $\varpi_0$ asymptotically approaches the fundamental mode of the isolated deformation $\epsilon V_p$, reflecting the effective decoupling between the main barrier and the deformation.

    Moreover, for the same stochastic deformation profile $V_p$ and strength $\epsilon$, the post-instability behavior exhibits nearly identical $\Im(\varpi_0)$--$|a|$ dependence in both PT and RW cases (see Fig.~\ref{figra}), including the same asymptotic frequency and comparable critical separation $a_c$. The only difference from the negative bump case is that, for stochastic perturbations, the critical separation follows $a_c \propto \epsilon^{-2}$ rather than $\epsilon^{-1}$. This remains consistent with the scaling structure described in Eq.~\eqref{eq1}. Moreover, this result confirms the conjecture raised in the previous section that $a_c$ depends not only on the deformation strength $\epsilon$ but also on the detailed form of $V_p$.

	So far, we have completed a systematic study of the effect of localized non-positive perturbations on BH stability from both the time domain and the frequency domain. We have clarified how these perturbations manifest in the frequency domain as the deformation moves from near the background potential to sufficiently close to the horizon. We have also discovered some commonalities of the effect of non-positive localized perturbations on BHs, in addition to the original spectral instability:
	\begin{itemize}
		\item[(1)] BH instability is not inevitable under perturbations; instead, it occurs only when the separation exceeds a critical distance $a_c$.
		
		\item[(2)] A purely imaginary mode $\varpi_0$ emerges and migrates along the imaginary axis toward larger imaginary values, eventually crossing into the upper half-plane and signaling instability.
		
		\item[(3)] $\varpi_0$ asymptotically approaches the fundamental mode of the isolated $\epsilon V_p$ as $a\to -\infty$.
		
		\item[(4)] The scaling of the critical distance depends on the deformation type: $a_c\propto \epsilon^{-1}$ for $\overline{V_p}<0$, and $a_c\propto \epsilon^{-2}$ for $\overline{V_p}=0$.
		
		\item[(5)] The critical distance $a_c$ is closely related to the shape of the deformation profile $V_p$.
	\end{itemize}
	Some of these commonalities will be verified in the theoretical derivation of the next section.
	\section{Theoretical Criterion for Instability}\label{sec5}
	As shown in Secs.~\ref{sec3} and~\ref{sec4}, the BH instability caused by negative and stochastic localized perturbations is not inevitable; instead, there exists a critical distance satisfying the relationship in Eq.~\eqref{eq1}. Only when the distance between the deformation and the main potential exceeds $|a_c|$ in the tortoise coordinate (i.e., when the perturbation is sufficiently close to the horizon) can the BH become unstable. Furthermore, we also observe in the QNM spectrum that negative and stochastic perturbations introduce a purely imaginary mode $\varpi_0$ carrying information about BH instability, and this mode will acquire a positive imaginary frequency as the system's ground mode when the BH becomes unstable. In this section, we will theoretically prove the above two findings by analogy with the Schr\"{o}dinger equation. Our proof contains following three steps.

    \subsection*{Step 1}
    In the first step, we will show that $\Im(\omega_0)<0$ will appear if the effective potential has non-positive mean-value.

	QNMs of BHs are determined by Eqs.~\eqref{KGe} and~\eqref{boundary}, whose complex frequencies $\{\omega_n\}$ form a discrete set. Since the potential $V$ is bounded from below, $\Im(\omega)$ is bounded from above, and the modes can be ordered as $\Im(\omega_n)\geq \Im(\omega_{n+1})$. The mode with the largest imaginary part, $\omega_0$, is referred to as the fundamental mode.
	
	The spectrum satisfies the symmetry
	\begin{equation}
		\omega = -\omega^*.
	\end{equation}
	Unless $\omega$ is purely imaginary, modes appear in pairs with identical imaginary parts, implying $\Im(\omega_0)=\Im(\omega_1)$. The late-time behavior is therefore governed by $\Im(\omega_0)$: $\Im(\omega_0)<0$ leads to exponential decay and the BH will be stable, while $\Im(\omega_0)>0$ implies exponential growth and instability.
	
	Eq.~\eqref{KGe} can be written in a Schr\"{o}dinger-like form by introducing
	\begin{equation}\label{ham}
		\hat{H} := -\frac{\td^2}{\td x^2} + V(x),
	\end{equation}
	where $\omega^2$ plays the role of an effective energy. Although the QNM problem is not self-adjoint, this analogy allows one to relate stability to the positivity of $\hat{H}$. Since $V(x)\to 0$ as $x\to\pm\infty$, negative values of $\omega^2$ correspond to normalizable bound-state-like solutions.

    For the fundamental QNMs of $\Im(\omega_0)>0$, the corresponding eigenfunctions are absolutely-integrable. As a result, the QNMs of $\Im(\omega_0)>0$ must have $\omega_0^2<0$. This can happen only if there exists an absolutely-integrable function $\phi(x)$ such that
    \begin{equation}\label{boundstate}
    	\int_{-\infty}^{+\infty}\phi^{*}\hat{H}\phi\,\td x < 0.
    \end{equation}
    Integrating by parts gives
    \begin{equation}\label{intHless0}
    	\int_{-\infty}^{+\infty}\phi^{*}\hat{H}\phi\,\td x
    	= \int_{-\infty}^{+\infty}\left[|\phi'|^2 + V(x)|\phi|^2\right]\td x
    	:= F_{V}[\phi].
    \end{equation}
    Therefore, if there exists a function $\phi$ satisfying
    \begin{equation}\label{FV}
    	F_V[\phi]<0,
    \end{equation}
    then the infimum of $F_V[\phi]$ over normalized functions is negative. Finding the solution of Eq.~\eqref{KGe} is equivalent to finding the function to minimize the function $F_V[\phi]$. When the infimum is attained, the minimizing function satisfies Eq.~\eqref{KGe} with $\omega_0^2$ equal to this minimum value, implying
    \begin{equation}
    	\omega_0^2<0.
    \end{equation}
    Consequently, the corresponding fundamental mode must be purely imaginary,
    \begin{equation}
    	\Re(\omega_0)=0.
    \end{equation}
    Such a spectral point lies on the imaginary axis and can satisfy $\Im(\omega_0)>0$, indicating BH instability. This provides a variational characterization of spectral points with $\Im(\omega_0)>0$ in terms of the functional $F_V[\phi]$. For the space $\mathcal{H}$ of all absolutely integrable $C^1$ functions on $\mathbb{R}$, spectral points with $\Im(\omega_0)>0$ exist precisely when the operator $\hat{H}$ in Eq.~\eqref{ham} is not semi-positive definite on $\mathcal{H}$. This establishes a direct connection between BH stability and the structure of the potential $V(x)$: if $V(x)\geq0$, then $F_V[\phi]\geq0$ for arbitrary $\phi$ and the system remains stable; while for sign-indefinite potentials, the sign of $\Im(\omega_0)$ depends on the detailed structure of $V(x)$.
	
	To derive a more explicit condition, we express the potential and test function in Fourier space:
	\begin{equation}\label{Fourier}
		V(x)=\int_{-\infty}^{+\infty}\mathcal{V}(\nu)e^{\ti\nu x}\,\td\nu,\quad
		\phi(x)=\int_{-\infty}^{+\infty}\varphi(\nu)e^{\ti\nu x}\,\td\nu.
	\end{equation}
	Substituting into $F_{V}$, we obtain
	\begin{equation}\label{FVFourier}
		F_{V}[\phi]=\int_{-\infty}^{+\infty}\td\nu\int_{-\infty}^{+\infty}\varphi^{*}(\nu)\,K(\nu,\tilde{\nu})\,\varphi(\tilde{\nu})\,\td\tilde{\nu},
	\end{equation}
	with kernel
	\begin{equation}\label{eqK}
		K(\nu,\tilde{\nu}):=\delta(\nu-\tilde{\nu})\,\nu\tilde{\nu}+\mathcal{V}(\nu-\tilde{\nu}).
	\end{equation}
	If the kernel $K(\nu,\tilde{\nu})$ is not semi-positive definite~\cite{Rachev2013}, then $F_{V}[\phi]$ can take negative values, implying the existence of modes with $\Im(\omega_0)$.
	
	Discretizing the integral, the kernel~\eqref{eqK} reduces to a Hermitian matrix $\mathbf{K}$ with elements
	\begin{equation}\label{Knl}
		\mathbf{K}_{nl}=\frac{\delta_{nl}}{\Delta\nu}\nu_{n}\nu_{l}+\mathcal{V}(\nu_{n}-\nu_{l}),
	\end{equation}
	where $\nu_{n}=n\Delta\nu$. A necessary condition for the semi-positivity of $\mathbf{K}$ is that the principal minor $\left( \begin{smallmatrix} \mathbf{K}_{00} & \mathbf{K}_{0n} \\ \mathbf{K}_{n0} & \mathbf{K}_{nn} \end{smallmatrix} \right)$ has a nonnegative determinant, which cannot be satisfied if $\mathbf{K}_{00}=\mathcal{V}(0)\leq0$. Since $\mathcal{V}(\nu)=(2\pi)^{-1}\int_{-\infty}^{\infty}V(x)\te^{-\ti\nu x}\td x$, we then see if
    \begin{equation}
    	\overline{V}:=\int_{-\infty}^{+\infty}V(x)\,\td x \leq 0,
    \end{equation}
    then the kernel $K(\nu,\tilde{\nu})$ cannot remain semi-positive definite, implying the existence of a mode with $\Im(\omega_0)>0$.

	This result provides a simple sufficient condition for instability and explains why perturbations with a non-positive spatial average tend to induce system instability, as observed in the numerical results. However, this conclusion cannot explain what we have found in above sections, since the original non-perturbed potential is positive and the total potential $V=V_b+\epsilon V_p$ has positive mean value.

    \subsection*{Step 2}	
	In order to explain what we have found in numerical calculations, we now consider $V(x)=V_{0}(x)+ \delta V(x-a)$, where $V_{0}(x)$ is positive potential and $\delta V(x)$ are localized absolutely integrable. We will show that if $\delta V(x)$ has non-positive mean-value, then there is a critical $a_c<0$ such that $\Im(\omega_0)>0$ if $a<a_c$.

    Assume that $\delta V(x)$ has non-positive mean-value, i.e. $\overline{\delta V}\leq0$. Based our conclusion in step 1, $\delta V(x)$ alone supports instability, i.e. there exists an absolutely-integrable function $\phi_0$ such that $F_{\delta V}[\phi_0]<0$.
	Construct the shifted function $\phi(x)=\phi_{0}(x-a)$. Then
	\begin{equation}\label{FV0}
		F_{V}[\phi]=F_{\delta V}[\phi_0]+\int_{-\infty}^{+\infty}V_{0}(x)|\phi_0(x-a)|^2\td x.
	\end{equation}
	Since $V_0(x)$ and $|\phi_0(x)|^2$ are integrable functions, the second term vanishes as $|a|\to\infty$ by the Riemann-Lebesgue lemma. Therefore,
	\begin{equation}\label{limFV}
		\lim_{a\to-\infty}F_{V}[\phi]=F_{\delta V}[\phi_0]<0.
	\end{equation}
	Thus, there must a critical $a_c<0$ such that $F_{\delta V}[\phi]<0$ when $a<a_c$, which implies the existence of $\Im(\omega_0)>0$.
    %\begin{equation}
%    	a_c := \inf \left\{ |a|\ge 0 \,\big|\, \exists\, \phi \ \text{such that } F_V[\phi]<0 \right\}.
%    \end{equation}
    The above analysis shows that if the isolated deformation $\delta V(x)$ alone can induce instability, then the combined potential
    \begin{equation}
    	V(x)=V_0(x)+\delta V(x-a)
    \end{equation}
    will also admit eigenfrequencies with fundamental $\Im(\omega_0)>0$ once the separation $a$ exceeds a critical value $a_c$.

    After using $\epsilon V_p$ to replace $\delta V$ in above proof, we now in fact have proven one main result in our numerically simulation: if a local perturbed potential $V_p$ is negative or stochastic with zero mean, then the foundmential mode $\omega_0$ will always satisfy $\Im(\omega_0)>0$ if $V_p$ is sufficiently closed to horizon.
	
    \subsection*{step 3}
    In this step, we will give the scaling relationship between $\epsilon$ and critical distance $a_c$, by which we give a proof for Eq.~\eqref{eq1}.

    Consider $\delta V=\epsilon W$ and the total  potential reads
	\begin{equation}\label{VeW}
		V(x)=V_{0}(x)+\epsilon W(x-a),
	\end{equation}
	with real $\epsilon$. For $\epsilon\overline{W}\leq0$, the instability criteria derived above implies that $\Im(\omega_0)>0$ once $a>a_c$. To extract the small-$\epsilon$ asymptotics of $a_c$, we write $\phi(x)=\phi_l(x)\te^{-\ti\omega x}$ and $\phi(x)=\phi_r(x)\te^{\ti\omega x}$ in the left and right regions, respectively, with boundary conditions $\phi_l(-\infty)=\phi_r(\infty)=1$. Here, the subscripts $l$ and $r$ refer to the left and right regions, respectively, while $\phi_l(x)$ and $\phi_r(x)$ are the corresponding wave-function coefficients after the asymptotic wave factors are extracted.
	
	At the critical distance $a=a_c$, the instability threshold is reached and the fundamental mode satisfies $\omega_0=0$. We then employ the shooting method and match the two solutions at a point $x_0\approx a_c/2$. The matching condition is equivalent to the vanishing of the Wronskian,
	\begin{equation}\label{Wron}
		\mathcal{W}[\phi_l,\phi_r](x_0)
		= \phi_l(x)\phi_r'(x)-\phi_l'(x)\phi_r(x)\big|_{x=x_0}=0.
	\end{equation}
	
	Since $V_0(x)$ and $W(x-a_c)$ are localized and well separated for large $a_c$, the solution $\phi_l$ is determined solely by $V_0(x)$, while $\phi_r$ is determined solely by $W(x-a_c)$.
	Under this approximation, Eq.~\eqref{KGe} reduces to
	\begin{equation}\label{newKG}
		\phi_{r}^{\prime\prime}-V_0(x)\phi_{r}=0,\quad
		\phi_{l}^{\prime\prime}-\epsilon W(x-a_c)\phi_{l}=0.
	\end{equation}
	
	Near the matching point $x_0\approx a_c$, both solutions can be approximated linearly as
    \begin{equation}
        \phi_{r}(x)=r_1+r_2 x,\quad
        \phi_{l}(x)=l_1+l_2 (x-a_c).
    \end{equation}
    Here, $r_1$ and $l_1$ are the constant coefficients of the linear expansions of $\phi_r$ and $\phi_l$, respectively, while $r_2$ and $l_2$ are the corresponding linear coefficients. Matching the two solutions and using Eq.~\eqref{Wron} yields
	\begin{equation}
		a_c=-\frac{r_1 l_2 - r_2 l_1}{r_2 l_2}.
	\end{equation}
	
	This relation holds for arbitrary $\epsilon$. In the small-$\epsilon$ limit, the leading-order solution of Eq.~\eqref{newKG} gives $\phi_l(x)\approx 1$, implying $l_1=1$ and $|l_2|\ll 1$. Therefore,
	\begin{equation}\label{ac}
		a_c=\frac{1}{l_2}+\mathcal{O}(1).
	\end{equation}
	A detailed calculation (see Appendix~\ref{AppA}) yields
	\begin{equation}\label{eq2}
		l_2=
		\begin{cases}
			\epsilon \overline{W}+\cdots, & \epsilon \overline{W}<0,\\
			-\epsilon^2 \overline{W_1^2}+\cdots, & \overline{W}=0,
		\end{cases}
	\end{equation}
	where $\overline{W_1^2}:=\int_{-\infty}^{+\infty} W_1^2(y)\,\td y$ and $W_1(x)=\int^{x}_{-\infty}W(y)\td y$. We then find
    \begin{equation}\label{eq2b}
		a_c=
		\begin{cases}
			\frac1{\epsilon \overline{W}}+\cdots, & \epsilon \overline{W}<0,\\
			-\frac1{\epsilon^2 \overline{W_1^2}}+\cdots, & \overline{W}=0,
		\end{cases}
	\end{equation}
	This result provides a theoretical explanation for the behavior observed in Ref.~\cite{mai_numerical_2026} and in Secs.~\ref{sec3} and~\ref{sec4}, as summarized in Eq.~\eqref{eq1}.
	
\begin{figure}
		\centering
		\includegraphics[width=0.66\textwidth]{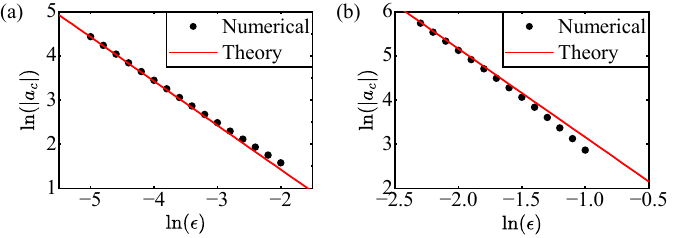}
		\caption{Critical distance $a_c$ as a function of deformation strength $\epsilon$. Panel (a) Localized negative Gaussian deformation. Panel (b) Stochastic deformation. Solid lines represent analytical results, while markers correspond to numerical simulations.}
		\label{figea}
	\end{figure}

    We compute the dependence of the critical distance $a_c$ on the deformation strength $\epsilon$ for the PT potential under both negative and stochastic perturbations. We consider the negative bump case with $V_p=-\te^{-(x-a)^2}$, and the stochastic perturbation case with $V_p=(x-a)\te^{-(x-a)^2}$. The corresponding theoretical predictions are given by $\ln(|a_c|)=-\ln(\epsilon)-\ln(\sqrt{\pi})$ and $\ln(|a_c|)=-2\ln(\epsilon)-\ln(\sqrt{2\pi}/8)$, respectively.  As shown in Fig.~\ref{figea}, the numerical data agree well with the theoretical prediction~\eqref{eq2b} for small $\epsilon$, while noticeable deviations appear as $\epsilon$ increases. This suggests that the deformation $\epsilon V_p$ can no longer be treated within the perturbative regime.
	
	Let us now make a summary on this section. The results derived above establish a complete theoretical framework for BH instability, encompassing both the existence conditions and their spectral properties. In particular, they demonstrate that BH instability is intrinsically tied to the non-positivity of the effective Hamiltonian, while the purely imaginary nature of eigenfrequencies with fundamental mode $\Im(\omega_0)>0$ emerges as a direct consequence of the associated negative directions of the operator. These results provide a unified explanation for the numerical observations in Secs.~\ref{sec3} and~\ref{sec4}, and highlight the fundamental role played by the structure of the deformation in triggering instability.

	\section{Conclusion And Discussion}\label{sec6}
	In this work, after verifying the instability of the model in the time domain, we have systematically investigated, within the frequency-domain, the effects of localized perturbations-arising from environmental effects as well as classical and quantum fluctuations-on the stability of BHs. Within a physically motivated framework, near-horizon fluctuations are modeled as localized deformations of the effective potential. We consider both deterministic negative perturbations and stochastic perturbations across a hierarchy of models, ranging from analytically tractable toy potentials (such as double-$\delta$, and PT potentials) to the physically relevant RW potential. A universal mechanism governing the onset of instability is identified.
	
	From a spectral perspective, non-positive perturbations affect the QNM spectrum through the emergence of a special purely imaginary mode. For small deformation displacements, this mode appears as a subdominant highly damped mode in the spectrum. As the deformation position is varied and gradually brought closer to the near-horizon region, the corresponding eigenfrequency moves continuously along the imaginary axis in the complex frequency plane, with its imaginary part monotonically increasing while the real part remains zero, leading to a progressive reduction of the decay rate. As a consequence, this mode crosses the original fundamental mode in the spectral ordering. When the deformation reaches a critical displacement, i.e. the perturbed potential is sufficiently closed to the horizon, the eigenfrequency enters the upper half of the complex plane and acquires a positive imaginary part, signaling the onset of instability. This provides a direct answer to the question: how non-positive perturbations modify the QNM spectrum in the frequency domain.
	
	Furthermore, within the frequency-domain, we confirm the scaling relation between the critical distance $|a_c|$ and the deformation strength proposed in Ref.~\cite{mai_numerical_2026}. In addition, in the limit corresponding to an infinitely near-horizon regime the fundamental mode asymptotically approaches that of the isolated deformation potential, indicating that the late-time dynamics is ultimately dominated by the localized deformation.

	On the theoretical side, we establish a rigorous criterion for BH instability, clarifying the precise conditions under which non-positive perturbations can trigger BH instability. By reformulating the problem in terms of the spectral properties of a Schr\"{o}dinger-like operator, we prove that eigenfrequencies with positive imaginary parts arise if and only if the corresponding effective Hamiltonian fails to be positive semidefinite. In particular, we derive a simple and physically transparent sufficient condition: if there is absolutely integrable function $\phi$ satisfies $F_V[\phi]<0$ (the $F_V[\phi]$ is defined in Eq.~\eqref{intHless0}) then a spectral point with $\Im(\omega)>0$ must exist. We further prove that all eigenfrequencies with $\Im(\omega)>0$ are necessarily purely imaginary. These results provide a unified theoretical explanation for the numerical observations and reveal that the instability originates from the global negativity of the effective potential. Meanwhile, we derive Eq.~\eqref{eq2b}, which provides a theoretical derivation of the power laws observed in Eq.~\eqref{eq1}, thereby answering whether these scaling behaviors can be derived theoretically.
	
	In summary, our results demonstrate that even localized perturbations with vanishing spatial average can destabilize BHs under appropriate conditions. This challenges the conventional expectation of the robustness of BH stability and highlights the potential role of near-horizon microscopic physics, such as quantum fluctuations or environmental disturbances, in influencing BH dynamics over long timescales.
	
	Finally, although eigenfrequencies with $\Im(\omega_0)>0$ are rigorously established in this work, their possibility to be found in the observations such as  gravitational-wave depends on if the requirements could be satisfied. If the deformation strength is really small then such instability will be hard to be found, since the timescale that such instability become considerable is $\mathcal{O}(1/\epsilon)$ or $\mathcal{O}(1/\epsilon^2)$. The necessary conditions for the instability are that: (1) the static negative or stochastic deformation on effective potential happens in the region that is sufficiently closed to event horizon, and (2) the deformation can keep for long enough time. Our current theoretical analyses support that such conditions do not completely impossible happen. We do not make any comment on if such conditions could be easy or difficult found in astronomic environments. Compared to the value of actual observations, our results are more meaningful in theory. It suggests that two BHs with almost identical metrics may have completely different stability if the slight difference between them occurs near the event horizon. It would be interesting to extend the present analysis to nonlocal or spatially extended perturbations, in order to explore how such global modifications compare with localized perturbations in triggering instability and to further understand their impact on BH stability.

    \backmatter
    \bmhead{Acknowledgements}
    	This work is supported by Natural Science Foundation of China under Grant No. 12375051, No. 12511540055 and Tianjin University Self-Innovation Fund Extreme Basic Research Project Grant No. 2025XJ22-0014 and No. 2025XJ21-0007. Z.-F. Mai is supported by  the "Hanjī" Action Plan (Guangxi Basic Research Program, Grand No.2026GXNSFBA00640240), the Guangxi Science and Technology Innovation Platform Program (Leitai Action Plan, Grant No.Guike LT2600640026) and the "Guangxi Highland of Innovation Talents'' Program.
    
    \begin{appendices}
    \section{QNM Calculation for \texorpdfstring{$\delta$}{delta}-Function Potentials}\label{App_delta}
    
    In this appendix, we present the detailed frequency-domain procedure used to calculate the QNMs for the $\delta$-function potentials. Since the potential vanishes everywhere except at the locations of the $\delta$ functions, the wave equation can be solved exactly in each interval between two successive $\delta$ functions. The propagation across the whole potential can therefore be expressed in terms of transfer matrices associated with the $\delta$ functions and free propagation between them.
    
    In a region where the potential vanishes, Eq.~\eqref{KGe} reduces to
    \begin{equation}
        \frac{\td^2\psi}{\td x^2}+\omega^2\psi=0.
    \end{equation}
    For a given complex frequency $\omega$, the general solution can be written as
    \begin{equation}
        \psi(x)
        =
        L\te^{-\ti\omega x}
        +
        R\te^{\ti\omega x},
    \end{equation}
    where $L$ and $R$ denote the amplitudes of the left- and right-moving waves, respectively. For convenience, we introduce the coefficient vector
    \begin{equation}
        \mathbf{C}
        =
        \begin{pmatrix}
            R\\
            L
        \end{pmatrix}.
    \end{equation}
    In the following, all transfer relations are defined from left to right. Thus, whenever the wave successively crosses a new potential or propagates through a free region, the corresponding matrix acts from the left on the existing coefficient vector.
    
    We first consider a single $\delta$-function potential of strength $H$ located at $x=0$,
    \begin{equation}
        V(x)=H\delta(x).
    \end{equation}
    Taking the position of the $\delta$ function as the local origin does not lead to any loss of generality, since the separation between different $\delta$ functions will be taken into account explicitly by the free-propagation matrix introduced below. The solutions immediately to the left and right of the $\delta$ function can be written as
    \begin{equation}
        \psi_-(x)
        =
        L_-\te^{-\ti\omega x}
        +
        R_-\te^{\ti\omega x},
    \end{equation}
    and
    \begin{equation}
        \psi_+(x)
        =
        L_+\te^{-\ti\omega x}
        +
        R_+\te^{\ti\omega x},
    \end{equation}
    respectively.
    
    The wave function is continuous across the $\delta$ function,
    \begin{equation}
        \psi_-(0)=\psi_+(0),
    \end{equation}
    which gives
    \begin{equation}
        L_-+R_-=L_++R_+.
        \label{delta_continuity}
    \end{equation}
    On the other hand, integrating
    \begin{equation}
        \frac{\td^2\psi}{\td x^2}
        +
        \left[\omega^2-H\delta(x)\right]\psi=0
    \end{equation}
    over an infinitesimal interval containing $x=0$ gives the jump condition
    \begin{equation}
        \psi'(0^+)-\psi'(0^-)=H\psi(0).
        \label{delta_jump}
    \end{equation}
    Since
    \begin{equation}
        \psi_\pm'(0)
        =
        -\ti\omega L_\pm+\ti\omega R_\pm,
    \end{equation}
    Eq.~\eqref{delta_jump} becomes
    \begin{equation}
        -\ti\omega L_+
        +\ti\omega R_+
        +\ti\omega L_-
        -\ti\omega R_-
        =
        H(L_-+R_-).
    \end{equation}
    Combining this equation with Eq.~\eqref{delta_continuity}, we obtain
    \begin{equation}
        R_+
        =
        \left(1-\frac{\ti H}{2\omega}\right)R_-
        -
        \frac{\ti H}{2\omega}L_-,
    \end{equation}
    and
    \begin{equation}
        L_+
        =
        \frac{\ti H}{2\omega}R_-
        +
        \left(1+\frac{\ti H}{2\omega}\right)L_-.
    \end{equation}
    Therefore, the coefficients on the two sides of the $\delta$ function are related by
    \begin{equation}
        \mathbf{C}_+
        =
        D(H,\omega)\mathbf{C}_-,
    \end{equation}
    where
    \begin{equation}
        D(H,\omega)
        =
        \begin{pmatrix}
            1-\dfrac{\ti H}{2\omega}
            &
            -\dfrac{\ti H}{2\omega}
            \\[3mm]
            \dfrac{\ti H}{2\omega}
            &
            1+\dfrac{\ti H}{2\omega}
        \end{pmatrix}.
        \label{delta_transfer}
    \end{equation}
    The matrix $D(H,\omega)$ thus gives the exact transformation of the wave coefficients across a single $\delta$ function.
    
    We next consider free propagation between two successive $\delta$ functions separated by a distance $d$. Since the potential vanishes in this region, propagation over the distance $d$ transforms the two components according to
    \begin{equation}
        R\rightarrow R\te^{\ti\omega d},
        \qquad
        L\rightarrow L\te^{-\ti\omega d}.
    \end{equation}
    The corresponding propagation matrix is therefore
    \begin{equation}
        \mathbf{C}(x+d)
        =
        P(d,\omega)\mathbf{C}(x),
    \end{equation}
    with
    \begin{equation}
        P(d,\omega)
        =
        \begin{pmatrix}
            \te^{\ti\omega d} & 0\\
            0 & \te^{-\ti\omega d}
        \end{pmatrix}.
        \label{delta_propagation}
    \end{equation}
    
    The above construction can be applied successively to an arbitrary finite number of $\delta$ functions. Consider $\delta$ functions located at
    \begin{equation}
        X_1<X_2<\cdots<X_N,
    \end{equation}
    with strengths $H_1,H_2,\ldots,H_N$. Starting from the left, crossing the first $\delta$ function gives
    \begin{equation}
        \mathbf{C}_1=D_1\mathbf{C}_0,
    \end{equation}
    where $D_j=D(H_j,\omega)$. Propagation from $X_1$ to $X_2$ then gives
    \begin{equation}
        \mathbf{C}_2=P_1D_1\mathbf{C}_0,
    \end{equation}
    where
    \begin{equation}
        P_j=P(X_{j+1}-X_j,\omega).
    \end{equation}
    After crossing the second $\delta$ function, one obtains
    \begin{equation}
        \mathbf{C}_3=D_2P_1D_1\mathbf{C}_0.
    \end{equation}
    Repeating this procedure, the coefficients on the two sides of the complete potential are related by
    \begin{equation}
        \mathbf{C}_{\rm end}
        =
        M(\omega)\mathbf{C}_0,
        \label{delta_total}
    \end{equation}
    where the total transfer matrix is
    \begin{equation}
        M(\omega)
        =
        D_NP_{N-1}D_{N-1}
        \cdots
        D_2P_1D_1.
        \label{delta_total_matrix}
    \end{equation}
    Thus, in accordance with the left-to-right propagation convention, each subsequent transformation is multiplied from the left.
    
    It remains to impose the QNM boundary conditions. With the time dependence $\te^{-\ti\omega t}$ adopted in this work, the QNM boundary conditions require a purely left-moving wave at the left asymptotic boundary and a purely right-moving wave at the right asymptotic boundary. In terms of the coefficient vector $\mathbf{C}=(R,L)^{\rm T}$, the coefficients at the left boundary can therefore be normalized as
    \begin{equation}
        \mathbf{C}_0
        =
        \begin{pmatrix}
            0\\
            1
        \end{pmatrix}.
    \end{equation}
    Writing the total transfer matrix as
    \begin{equation}
        M(\omega)
        =
        \begin{pmatrix}
            M_{11}(\omega) & M_{12}(\omega)\\
            M_{21}(\omega) & M_{22}(\omega)
        \end{pmatrix},
    \end{equation}
    Eq.~\eqref{delta_total} gives
    \begin{equation}
        \mathbf{C}_{\rm end}
        =
        \begin{pmatrix}
            M_{12}(\omega)\\
            M_{22}(\omega)
        \end{pmatrix}.
    \end{equation}
    At the right asymptotic boundary, the QNM boundary condition requires the left-moving component to vanish, so that
    \begin{equation}
        \mathbf{C}_{\rm end}
        \propto
        \begin{pmatrix}
            1\\
            0
        \end{pmatrix}.
    \end{equation}
    Consequently, the QNM frequencies are determined by
    \begin{equation}
        M_{22}(\omega)=0.
        \label{delta_qnm}
    \end{equation}
    The QNM problem is therefore reduced to finding the isolated zeros of $M_{22}(\omega)$ in the complex-frequency plane. Since both $D_j$ and $P_j$ are known analytically, $M_{22}(\omega)$ can be evaluated directly for any trial complex frequency, and its complex roots give the QNM spectrum.
    
    We finally note that the transfer-matrix construction above differs in implementation from the direct integration and matching procedure commonly used for smooth effective potentials. In the latter case, two solutions satisfying the QNM boundary conditions at the left and right asymptotic boundaries are usually integrated numerically toward a matching point, typically chosen near the peak of the main potential barrier, and the QNM frequencies are determined by requiring the two solutions to be linearly dependent, or equivalently by requiring their Wronskian to vanish. For $\delta$-function potentials, however, the wave equation is exactly solvable between successive $\delta$ functions, while the matching conditions across each $\delta$ function are also known analytically. The propagation and matching processes can therefore be performed exactly through a finite product of matrices, without numerical integration in the free regions. The transfer-matrix construction can thus be regarded as an analytic implementation of the usual direct integration and matching procedure for this special class of potentials: both solve the same eigenvalue problem subject to the same QNM boundary conditions, while the transfer-matrix form avoids unnecessary numerical integration when the piecewise solutions and matching conditions are available analytically.
        
    \section{Perturbed P\"oschl--Teller Potential}\label{App_pt}
    	The perturbed form of Eq.~\eqref{KGe} can be written as
    	\begin{equation}\label{KGp}
    		\left(
    		-\frac{\td^2}{\td x^2}
    		-\omega^2
    		+V_b(x)
    		+\epsilon V_p(x)
    		\right)\psi(x)=0,
    	\end{equation}
    	where $V_b(x)$ denotes the background PT potential and $V_p(x)$ represents the localized deformation.
    	
    	In the absence of the deformation ($\epsilon=0$), the main potential barrier reduces to
    	\begin{equation}
    		V_b(x)=\frac{1}{\cosh^2 (x)},
    	\end{equation}
    	for which two independent solutions adapted to the QNM boundary conditions can be written as~\cite{PhysRevX.11.031003}
    	\begin{equation}\label{hyperc}
    		\psi_{0}^{\pm}(x)
    		=
    		\te^{\ti\omega \ln(\cosh x)}
    		\,{}_2F_1
    		\left(\alpha,\beta;\gamma;z^\pm\right).
    	\end{equation}
    	where $\alpha=(1-\ti\sqrt{3})/2-\ti\omega$, $\beta=(1+\ti\sqrt{3})/2-\ti\omega$, $\gamma=1-\ti\omega$, and $z^{\pm}=1/(1+\te^{\pm2x})$. The solutions $\psi_0^{+}$ and $\psi_0^{-}$ asymptotically behave as $\psi_0^{+}\sim\te^{+\ti\omega x}$ as $x\to+\infty$ and $\psi_0^{-}\sim\te^{-\ti\omega x}$ as $x\to-\infty$, satisfying the QNM boundary conditions at the right and left asymptotic boundaries, respectively.
    	
    	Assuming that the deformation strength $\epsilon$ is sufficiently small, the solution of Eq.~\eqref{KGp} can be written in the form
    	\begin{equation}
    		\psi(x)=\psi_0(x)\left(1-\frac{1}{2}U(x)F(x)\right),
    	\end{equation}
    	where $F(x)$ encodes the deformation induced by the localized deformation and $U(x)$ is a prescribed auxiliary function introduced to simplify the resulting equation.
    	
    	Substituting this ansatz into Eq.~\eqref{KGp} and expanding to linear order in $\epsilon$, one obtains a second-order differential equation for $F(x)$ of the form
    	\begin{equation}\label{eqF}
    		\frac{\td^2 F}{\td x^2}
    		+ A(x)\frac{\td F}{\td x}
    		+ B(x)F+S(x)=0,
    	\end{equation}
    	where the coefficient functions read
    	\begin{equation}\label{eqABS}
    		\begin{cases}
    			A(x)=\frac{2\psi_0^\prime (x)}{\psi_0(x)}+\frac{2U^\prime(x)}{U(x)},\\
                \\
    			B(x)=2\frac{\psi_0^\prime(x) U^\prime(x)}{\psi_0(x)U(x)}+\frac{U^{\prime\prime}(x)}{U(x)}-\epsilon V_p(x),\\
                \\
    			S(x)=\frac{2V_p(x)}{U(x)},\\
    		\end{cases}
    	\end{equation}
    	here the prime denotes differentiation with respect to $x$. To ensure the solvability and convergence of the above equation, we impose the following constraints on the function $U(x)$:
    	\begin{itemize}
    		\item[(1)] $U(x)\big|_{x\to\pm\infty}=0$, which guarantees that the asymptotic boundary conditions required for QNM calculations are satisfied at spatial infinity;
    		\item[(2)]
    		The function $U(x)$ must decay sufficiently rapidly at spatial infinity to suppress exponentially growing numerical contamination during the integration procedure;
    		\item[(3)]The asymptotic decay of $U(x)$ should not exceed that of $V_p(x)$, such that the ratio $2V_p/U$ remains finite as $x\to\pm\infty$.
    	\end{itemize}
    	
    	For the Gaussian bump deformation
    	\begin{equation}
    		V_p(x)=\te^{-(x+a)^2},
    	\end{equation}
    	we consider the left asymptotic region ($x<0$) and choose
    	\begin{equation}
    		U(x)=\frac{\sqrt{\pi}}{2}\left[\erf(x+a)+1\right],
    	\end{equation}
    	which satisfies
    	\begin{equation}
    		\frac{\td U(x)}{\td x}
    		=
    		\te^{-(x+a)^2}.
    	\end{equation}
    	This choice satisfies all the asymptotic conditions listed above. Substituting the above choice of $U(x)$ into Eq.~\eqref{eqABS}, one obtains
    	\begin{equation}\label{eqABS2}
    		\begin{cases}
    			A(x)=\frac{2\psi_0^{\prime}(x)}{\psi_0(x)}+\frac{4\te^{-(x+a)^2}}{\sqrt{\pi}(\erf(x+a)+1)},\\
                \\
    			B(x)=\frac{4\te^{-(x+a)^2}}{\sqrt{\pi}(\erf(x+a)+1)}\left(\frac{\psi_0^{\prime}(x)}{\psi_0(x)}-(x+a)\right)-\epsilon V_p(x),\\
                \\
    			S(x)=\frac{4\te^{-(x+a)^2}}{\sqrt{\pi}(\erf(x+a)+1)},\\
    		\end{cases}
    	\end{equation}
    	For convenience, we further introduce the auxiliary function
    	\begin{equation}\label{eqh}
    		h(x)=-\frac{\sqrt{\pi}(\erf(x+a)+1)}{\te^{-(x+a)^2}}.
    	\end{equation}
    	Eq.~\eqref{eqABS2} can be recast into the form
    	\begin{equation}\label{eqABS3}
    		\begin{cases}
    			A(x)=\frac{2\psi_0^{\prime}(x)}{\psi_0(x)}-\frac{4}{h(x)},\\
                \\
    			B(x)=-\frac{4}{h(x)}\left(\frac{\psi_0^{\prime}(x)}{\psi_0(x)}-(x+a)\right)-\epsilon V_p(x),\\
                \\
    			S(x)=-\frac{4}{h(x)},\\
    		\end{cases}
    	\end{equation}
    	In the asymptotic region $x\to-\infty$, the background solution behaves as
    	\begin{equation}
    		\psi_0^{-}(x)\sim \te^{-\ti\omega x},
    	\end{equation}
    	while the deformation satisfies $V_p(x)\to0$. Using these asymptotic behaviors, we expand $F(x)$ as a Laurent series in powers of $(x+a)^{-1}$,
    	\begin{equation}
    		F(x)=\sum_{n=0}^{\infty}\frac{c_n^-}{(x+a)^n},
    	\end{equation}
    	and keeping terms up to order $(x+a)^{-3}$, one obtains
    	\begin{equation}
    		c_{0}^{-}=0,\qquad
    		c_{1}^{-}=1,\qquad
    		c_{2}^{-}=-\ti\omega,\qquad
    		c_{3}^{-}=-\omega^2-1.
    	\end{equation}
    	
    	An analogous construction can be carried out in the right asymptotic region ($x>0$), where
    	\begin{equation}
    		U(x)=\frac{\sqrt{\pi}}{2}\left[\erf(x+a)-1\right].
    	\end{equation}
    	The corresponding expansion coefficients are
    	\begin{equation}
    		c_{0}^{+}=0,\qquad
    		c_{1}^{+}=1,\qquad
    		c_{2}^{+}=\ti\omega,\qquad
    		c_{3}^{+}=-\omega^2-1.
    	\end{equation}
    	
    	Finally, the corrected wavefunctions
    	\begin{equation}
    		\psi^{\pm}(x)
    		=
    		\psi_{0}^{\pm}(x)
    		\left(
    		1-\frac{\epsilon U^{\pm}(x)F^{\pm}(x)}{2}
    		\right)
    	\end{equation}
    	are used as the left and right asymptotic solutions. The QNM frequencies of the PT potential perturbed by a localized negative Gaussian bump can then be obtained through the standard shooting and matching procedure. This construction provides the initial data required for the numerical determination of the perturbed QNM spectrum.
    \section{Perturbed Regge--Wheeler Potential}\label{App_rw}
    	
    	In this appendix, we briefly summarize how the perturbative construction developed in Appendix~\ref{App_pt} can be combined with the analytic continuation prescription of Ref.~\cite{Plamen_P_Fiziev_2007} to determine the QNM spectrum of the perturbed RW potential. We do not repeat the derivation of the analytic continuation method itself, and only outline the ingredients relevant for the present work.
    	
    	As shown in Ref.~\cite{Plamen_P_Fiziev_2007}, the RW equation admits two independent local solutions near the event horizon,
    	\begin{equation}
    			\psi_{0}^{\pm}(r)
    			=\,r^{s+1}
    			\te^{-\ti \omega r\mp\ti\omega\ln(1-r)}\times
    			\mathrm{HeunC}
    			\left(
    			-2\ti\omega,
    			\mp2\ti\omega,
    			2s,
    			2\omega^2,
    			\alpha,
    			1-r
    			\right),
    	\end{equation}
    	where $\mathrm{HeunC}$ denotes the confluent Heun function and
    	\begin{equation}
    		\alpha = 2\omega^2+s^2-\ell(\ell+1).
    	\end{equation}
    	The parameters $\ell$ and $s$ represent the multipole number and the spin of the perturbation field, respectively. The solutions $\psi_0^{-}$ and $\psi_0^{+}$ describe the ingoing and outgoing waves at the event horizon. Details on the numerical implementation of the confluent Heun function can be found in Ref.~\cite{8553032}.
    	
    	Although an exact outgoing-wave solution at spatial infinity is unavailable, Ref.~\cite{Plamen_P_Fiziev_2007} proposed an equivalent QNM prescription based on analytic continuation in the complex-$r$ plane. Specifically, one considers the asymptotic direction
    	\begin{equation}
    		r=|r|\te^{-\ti\left(\frac{\pi}{2}-\arg(\omega)\right)},
    		\qquad |r|\to+\infty,
    	\end{equation}
    	along which
    	\begin{equation}
    		\lim_{|r|\to+\infty}
    		\left|
    		r^{s+1}
    		\mathrm{HeunC}
    		\left(
    		-2\ti\omega,
    		\mp2\ti\omega,
    		2s,
    		2\omega^2,
    		\alpha,
    		1-r
    		\right)
    		\right|
    		=0.
    	\end{equation}
    	Under this analytic continuation, the unwanted asymptotic component becomes exponentially suppressed, allowing the QNM boundary condition to be imposed through the vanishing of its coefficient.
    	
    	The radial coordinate $r$ and the tortoise coordinate $x$ are related by
    	\begin{equation}
    		\td x=\frac{\td r}{f(r)}.
    	\end{equation}
    	For a Schwarzschild BH with horizon radius $r_h=1$, one obtains
    	\begin{equation}
    		x(r)=r+\ln(r-1)+C,
    	\end{equation}
    	whose inverse relation is
    	\begin{equation}
    		r(x)=1+W\left(\te^{x-1-C}\right),
    	\end{equation}
    	where $W(x)$ denotes the Lambert-$W$ function. The integration constant
    	\begin{equation}
    		C=-r_{\mathrm{peak}}-\ln(r_{\mathrm{peak}}-1)
    	\end{equation}
    	is chosen such that the RW potential reaches its maximum at $x=0$.
    	
    	Since the perturbative construction in Appendix~\ref{App_pt} relies only on the existence of asymptotic ingoing and outgoing solutions together with the localized nature of the deformation, the same strategy can be generalized to the RW potential through the above coordinate transformation.
    	
    	Starting from the corrected near-horizon solution $\psi^{-}(r)$ as the initial condition, the perturbed RW equation is numerically integrated from the near-horizon region ($x<-a$) toward the asymptotic region ($x>-a$). The resulting solution can then be decomposed as
    	\begin{equation}\label{c9}
    		\psi(r)
    		=
    		C^{+}\psi^{+}(r)
    		+
    		C^{-}\psi^{-}(r),
    	\end{equation}
    	where $C^{+}$ and $C^{-}$ denote the outgoing and ingoing wave amplitudes, respectively.
    	
    	The QNM boundary condition is imposed through
    	\begin{equation}
    		\lim_{|r|\to+\infty}
    		\left|
    		\psi\left(
    		|r|\te^{-\ti(\pi/2-\arg(\omega))}
    		\right)
    		\right|
    		=0,
    	\end{equation}
    	which suppresses the unwanted asymptotic component along the complex contour. The QNM frequencies are then determined by requiring
    	\begin{equation}
    		C^{-}=0.
    	\end{equation}
    
        We generally assume that the effect of $V_p$ can be neglected for $|x-a|>15$ (we have fixed $r_h=1$), which allows us to truncate the integration when decomposing Eq.~\eqref{c9}. When $a\ll 0$, the cutoff in tortoise coordinates satisfies $x_{\text{cut-off}}\ll 0$ ($r_{\text{cut-off}}\approx 1$). However, when $a$ approaches the RW potential, the cutoff may cross the RW barrier and lie on its right-hand side. Since our numerical code for evaluating the HeunC function cannot yield reliable results for $r\gtrapprox 7$, we are restricted to computations with $a\lessapprox -7$. Even if $V_p$ remains negligible for smaller cutoff distances, it remains unclear whether QNM spectra with smaller $|a|$ can be reliably determined. Echo effects always exist in the system, and their occurrence frequency increases as $|a|$ decreases. For sufficiently small $|a|$, frequent echo signals contaminate the fundamental mode approximation extracted from time-domain evolution, while the system rapidly enters the power-law tail regime. This prevents us from validating the solutions obtained in the frequency domain. For these reasons, we compute QNM spectra only for $a< -7$.

        \section{Asymptotic evaluation of \texorpdfstring{$l_2$}{l2}}\label{AppA}
    	In this appendix, we derive the asymptotic behavior of $l_2$ in the small-$\epsilon$ limit.
    	
    	Starting from Eq.~\eqref{newKG}, we introduce the shifted coordinate
    	\begin{equation}
    		z = x - a_c,
    	\end{equation}
    	so that the equation for $\phi_l$ becomes
    	\begin{equation}
    		\phi_l''(z) - \epsilon W(z)\phi_l(z)=0.
    	\end{equation}
    	
    	This equation can be rewritten as the integral equation
    	\begin{equation}\label{int_eq}
    		\phi_l(z)=1-\epsilon\int^{z}_{-\infty}(y-z)W(y)\phi_l(y)\,\td y.
    	\end{equation}
    	
    	To facilitate the expansion, we define the auxiliary functions
    	\begin{equation}
    		W_1(x)=\int^{x}_{-\infty}W(y)\,\td y,\quad
    		W_2(x)=\int^{x}_{-\infty}(y-x)W(y)\,\td y,
    	\end{equation}
    	which satisfy
    	\begin{equation}
    		W_1(\infty)=\overline{W},\quad
    		W_1'=W,\quad
    		W_2'=-W_1,\quad
    		W_2(-\infty)=0.
    	\end{equation}
    	
    	The integral equation~\eqref{int_eq} can be solved perturbatively. Up to second order in $\epsilon$, one obtains
    	\begin{equation}
    		\phi_l(z)=1-\epsilon W_2(z)+\epsilon^2\int^{z}_{-\infty}(y-z)W(y)W_2(y)\,\td y+\mathcal{O}(\epsilon^3).
    	\end{equation}
    	
    	The coefficient $l_2$ is determined by
    	\begin{equation}
    		l_2 = \phi_l'(z)\big|_{z\to \infty}.
    	\end{equation}
    	Evaluating the derivative yields
    	\begin{equation}
    		l_2 = \epsilon \overline{W} -\epsilon^2 \int_{-\infty}^{\infty} W_1'(y)W_2(y)\,\td y + \mathcal{O}(\epsilon^3).
    	\end{equation}
    	For $\overline{W}\neq 0$, the leading term gives
    	\begin{equation}
    		l_2 = \epsilon \overline{W}+\mathcal{O}(\epsilon^2).
    	\end{equation}
    	If $\overline{W}=0$, integration by parts gives
    	\begin{equation}
    		l_2 =- \epsilon^2 \int_{-\infty}^{\infty} W_1^2(y)\,\td y + \mathcal{O}(\epsilon^3).
    	\end{equation}

    \end{appendices}
    \bibliography{paper2}
\end{document}